# Trusted Certificates in Quantum Cryptography

William Perkins

## Abstract

This paper analyzes the performance of Kak's three stage quantum cryptographic protocol based on public key cryptography against a man-in-the-middle attack. A method for protecting against such an attack is presented using certificates distributed by a trusted third party.

## Introduction

The three-stage quantum cryptographic protocol introduced in [1] uses properties of quantum mechanics to apply a series of secret commutative transformations on a qubit which is exchanged over a public channel. Unlike the BB84 protocol [2], all three transmissions are performed on quantum channels, eliminating the possibility of successful eavesdropping at any point during the exchange.

The protocol is, however, susceptible to a man-in-the-middle attack [3], in which the adversary (Eve) intercepts a message from Alice and creates a new message to send to Bob (Fig.1). Eve performs the exchange with Alice using the original message, while Bob performs the exchange using the newly created message. At the final stage, Eve has the original message in decrypted form, while Bob has the message created by Eve. It is unknown to Bob that his message originated from Eve and not Alice.

One approach used in classical cryptography to defend against the man-in-the-middle attack is authentication through the use of trusted certificates [4]. The same concept can be extended to quantum cryptographic protocols as well. This paper presents an alteration of Kak's protocol which includes the use of quantum entangled certificates, distributed by a trusted authority, for the purpose of authentication. The strength in this modification relies on the inability of Eve to duplicate and existing entanglement.

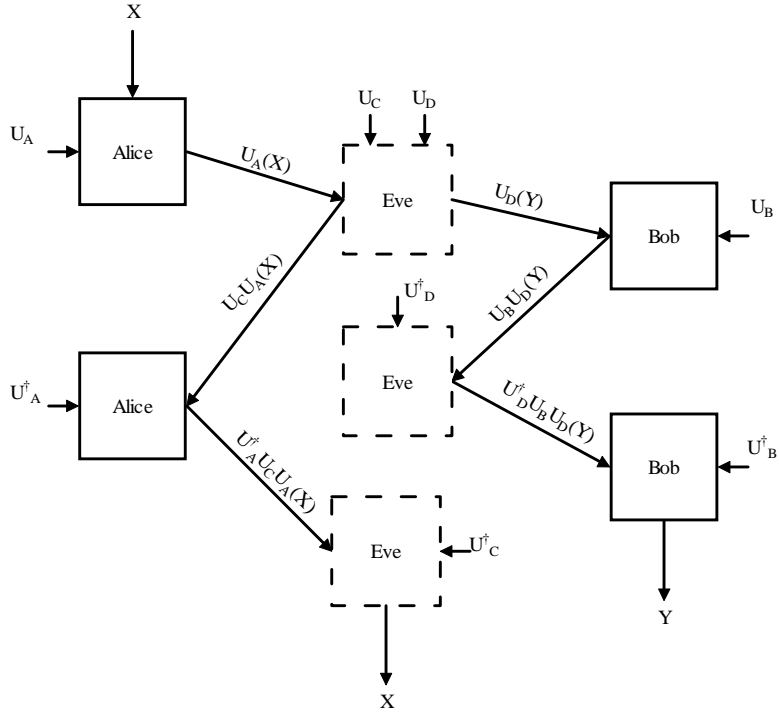

**Figure1**. Man-in-the-middle attack

## Authentication

Let a certificate authority, CA, distribute a set of entangled qubits [5] [6] to Alice and Bob. The sequence of entangled pairs constitutes Bell states which are not known to Eve. The pre-shared master key between the clients and the CA consists of the type of Bell states that are transmitted as session certificates.

$$B_{AB} = [\beta_{0i1} \quad \beta_{0i2} \quad \cdots \quad \beta_{0iN}] \text{ where } K_{CA} = [i1 \quad i2 \quad \cdots \quad iN] \in \{0,1\} \text{ is the master key}$$

$$\beta_{00} = \frac{|00\rangle + |11\rangle}{\sqrt{2}} \quad \beta_{01} = \frac{|01\rangle + |10\rangle}{\sqrt{2}}$$

When Alice wants to send a message to Bob, she appends her certificate to the message. The message exchange is carried out just as demonstrated previously with only one minor change. At the end of the exchange, Bob collapses his certificate and compares it to Alice's as a method of authentication (Fig. 2). If Bob's newly received message is truly authentic, then the included certificate will correspond to Bob's certificate. As a further enhancement, the placement of the certificate within the message can be a function of the master key.

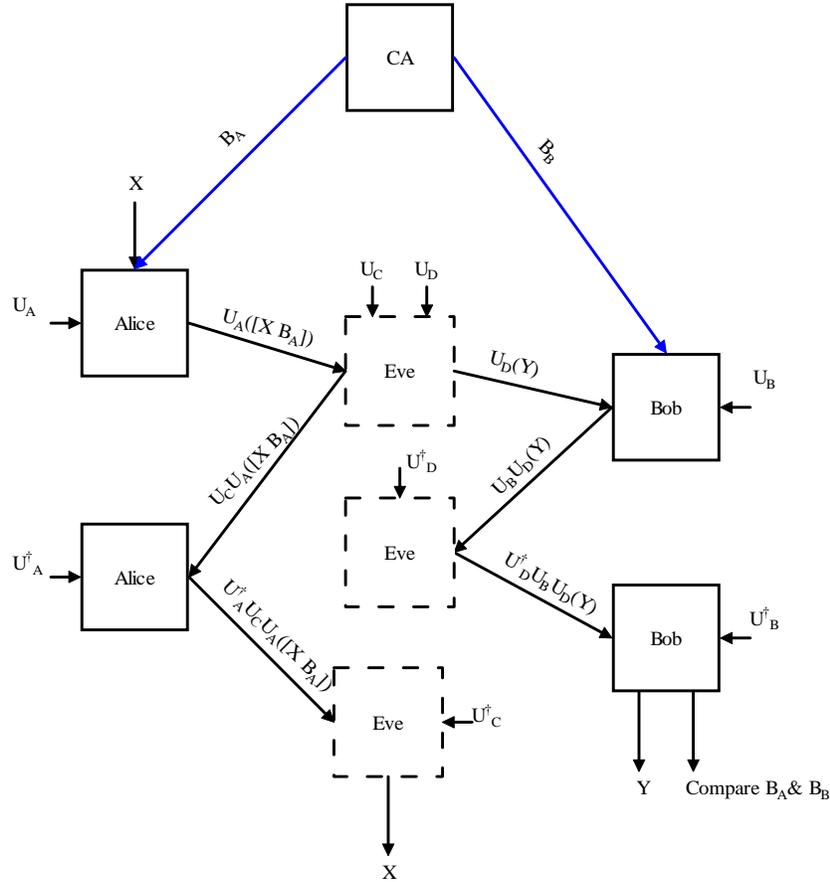

**Figure 2.** The new protocol

### Example 1

Suppose the master key (known by Alice and Bob) indicates that the type of Bell states transmitted as the certificate will be $B_{AB} = [\beta_{01} \quad \beta_{00} \quad \beta_{00} \quad \beta_{01}]$. For Bob to check the authenticity of the message received, he simply needs to perform an XOR of his certificate, the received certificate, and the master key.

|   |   |   |   |   |
|---|---|---|---|---|
|   | 1 0 0 1 | Master Key |
| $\oplus$ | 0 1 0 1 | Alice's Certificate |
|   | 1 1 0 0 | Bob's Certificate |
|   | 0 0 0 0 |   |

**Authentic**

|   |   |   |   |   |
|---|---|---|---|---|
|   | 1 0 0 1 | Master Key |
| $\oplus$ | 0 0 0 1 | Alice's Certificate |
|   | 1 1 0 0 | Bob's Certificate |
|   | 0 1 0 0 |   |

**Not Authentic**

If the result of the XOR operation contains a 1, then the certificate is not authentic.

## Conclusion

This paper extends the use of certificates of authenticity from classical cryptography to quantum cryptography. A method using entangled qubits distributed by a trusted third party provides a significant defense against the man-in-the-middle attack to which other protocols including BB84 are vulnerable. Recent advances in technologies for creating quantum entangled particles [6] confirm the feasibility of the realization of such a protocol.